# AN OPERATING SYSTEM FOR EXTRA LONG URBAN TRAINS

Carlos F. Daganzo

Institute of Transportation Studies

University of California, Berkeley, California, 94720, USA

(July 16, 2021)

## ABSTRACT

An operating system (*OS*) for subways and other urban railways is presented. The system uses extra-long trains (*XLT*s) that can protrude beyond both ends of the station platforms. No added infrastructure is needed--only more rolling stock. The system's only technological requirement is that the doors in different parts of each train (e.g., its cars) can be operated independently. The system can preserve the level of service and evenly fill with passengers all the cars of an *XLT* so no space is wasted. With sufficiently long trains, a railway's productivity can be more than doubled.

The proposed *OS* has a train-side and a passenger-side. On the train side, it includes train organization and station-stopping protocols and on the passenger side a new information system that organizes passengers at the platforms as required by the train-side protocols. These protocols specify the composition of each train and what it does at each station; i.e., whether it stops or not; how it aligns its doors along the platform; whether each door opens or not; and the set of destinations advertised by each open door. A general menu of train-side protocols is presented, as well as a mathematical framework for their optimization and analysis. Numerous examples illustrating key concepts are also presented.

# 1. INTRODUCTION

Subway systems in some of the world's megacities are overcrowded. See e.g., Daganzo (2019) for examples and discussion. Although overcrowding can obviously be eliminated with longer trains, the lengths of existing station platforms have constrained what can be done. This is a problem because expanding stations is expensive. It therefore seems important to see if *XLT*s (trains substantially longer than the platforms) can be introduced, while still filling them with passengers.

This important question appears to have been first addressed in Villalobos and Munoz (2014), which introduces the concept of "*Hypertren*". This document proposes two clever strategies for using trains longer than the station platforms. The first strategy doubles the train length and allows all passengers to board anywhere along the platform. The second strategy directs some of the passengers to either the front or the rear of the platform but uses shorter trains. Both strategies, however, have three drawbacks: (i) they cannot be efficiently used for transfers; (ii) they cannot always be applied in both directions of two-way lines; and (iii) they may require the expansion of end-of-line stations.

A slight but useful variant of the second, shorter-train strategy seems to have been independently proposed in Avedisian (2017). The variant is appealing because it does not require that the passengers be organized at the platforms with an information system. On the other hand, this very simplification has another drawback: (iv) it prevents all cars from being filled to capacity when the train flows through a line's maximum load point.

A similar idea to the first, longer-train, *Hypertren* strategy in Villalobos and Munoz (2014), and still suffering from some of the drawbacks mentioned above, has also been recently proposed (Daganzo, 2018). The idea has been piloted in Beijing, China (Zhou, 2020).

This paper adds to this emerging body of knowledge by presenting a general operating system for extra-long trains (*OS-XLT*) that overcomes the aforementioned drawbacks while allowing for even longer trains.[1] Unlike previous *XLT* strategies, *OS-XLT* can direct passengers to underused sections of a train and in this way control each train's load profile as it travels along its route—even in bidirectional lines. This alleviates drawbacks (ii) and (iv) as trains can then be filled evenly with passengers at the maximum load points to improve comfort and capacity. The paper will also show how the other two drawbacks are overcome, and how the length of ordinary trains can be tripled.

The new system has two components: (a) a new passenger information system (*PAXIS-XLT* for short), and (b) a new train operation system (*TOPS-XLT*) with protocols for configuring trains and specifying what they do at each station. In the system's most ambitious form, the station platforms are reconfigured with "smart" boarding gates.

---

[1] The generalization is encapsulated in three preliminary patents (Daganzo, 2020). The patents have lapsed so the ideas are now in the public domain. This paper formalizes their presentation into said domain.

The paper also describes a mathematical programming framework to help choose the best system configuration for any particular application. To bring it all together, the paper is organized as follows. Section 2 below, describes the passenger information system, Section 3 the train-side protocols, Sections 4 and 5 some specific instances, and Sections 6 and 7 the optimization framework. Section 8 closes with some caveats and suggestions for further work.

## 2. THE PASSENGER INFORMATION SYSTEM

After introducing some terminology in Section 2.1, the narrative below describes in Sections 2.2 to 2.5 the elements of *PAXIS-XLT* by location: outside the stations; in the stations; at the platforms and in the trains. Section 2.6, then explains how these elements work together.

### 2.1 Definitions

<u>Train</u>: A multi-unit vehicle that stops for embarkation and disembarkation at station platforms making a run between two end-of-line stations without changing its composition. A train has external doors that allow passengers to embark and disembark at the stations visited.

<u>Platform</u>: The part of a station facing the trains where passengers embark and disembark. A station may have multiple platforms. A platform may serve multiple train routes.

<u>Gate (or platform gate)</u>: That part of a platform that faces a train door. Passengers must pass through a gate/door combination to embark or disembark. Gates can include physical structures, or be purely virtual; i.e., with signs only. The only requirements are that they should be readily identifiable by passengers, and be always aligned with the doors of stopped trains.[2]

<u>Units (or train units)</u>: The smallest continuous parts of a train whose doors can be operated independently; e.g., a car. A train's units must partition the train. Each unit has a distinct set of doors. For each train, units are characterized by their length and their passenger holding capacity.

### 2.2 Information outside the stations

The purpose of information outside the stations is to allow passengers to decide whether to use the station. At a minimum, this information should include a list with the destinations served by the station, perhaps complemented by maps. Optionally, for each destination *PAXIS-XLT*

---

[2] If the platform serves a unique type of train with evenly spaced doors then the gates provided at the platform can be placed at fixed locations, with their number equal to the maximum number of doors that a stopped train can align with the platform. This is common practice. However, if the platform serves multiple train types with different or unevenly spaced doors, this is not possible. In this case, an agency using *OS-XLT* can provide many

can display the expected departure time of the next train (or trains) to serve it. The way in which the information is conveyed is up to the subway agency. For example, if destinations can be grouped into a few classes that always share the same departure time, *PAXIS-XLT* could display the departure times of the different classes. The foregoing is common practice, so it is not described further.

## 2.3 Information inside the station.

The purpose of this information is to tell passengers bound for specific destinations which platform to access—and optionally the next departure time. If desired, this information can also be given for the next few trains. Again, the provision of this information is common practice so it is not described further.

## 2.4 Embarkation information at platforms

The purpose of this information is to tell passengers bound for specific destinations at what gate to wait for embarkation in the next train serving the destination. Ideally, the information should be provided at the platform's entrances, throughout the platform and at the gates themselves.

The information should be changeable so the set of *gates for a given destination can be varied across successive trains*. This feature is believed to be new for high-frequency subway systems. Note that it is possible to display for any given train a different set of destinations at every single gate; and even to *display no destinations at all at some of the gates through which some passengers will be disembarking*. This fine-grained flexibility opens the door for sophisticated protocols as Section 4.2 shall illustrate.

A possible way of conveying the desired gate-destination information is with variable message signs at each gate that indicate the destinations served; e.g., as is done at airports. The information could be displayed in many ways; e.g., by a list of destinations served; or more intuitively by a map displaying the route served by the next train with the destinations served at the gate highlighted, e.g. with flashing lights. If some gates are to be used for disembarkation only, an embarkation ban can also be physically enforced; e.g., with barriers.

Optionally, each gate would also include similar information for the next few arriving trains.

## 2.5 Disembarkation information inside the train

The purpose of this information is to tell passengers where and how to disembark. This can be accomplished with displays next to each door that indicate the destinations that will be served. This is commonly done so it does not need much description.

---

closely spaced virtual gates all along the platform with the idea of activating for each train only the gates that align with its doors.

The door displays must be consistent with the gate displays of the stations already visited. So passengers already on-board can be reassured that they will be able to disembark where they intended. What is different now is that *displays near each door may be different from one another*. Also different is that the *displays can add new destination stations while a train is in motion*.

The features of Sections 2.4 and 2.5 allow operators to reserve some train units for passengers with certain O-Ds; and to do this adaptively, e.g., based on the observed passenger loads. The utility of all the features is that passengers with different O-D pairs can be directed to specific sets of train units. This capability can be exploited by smart train operation protocols to maximize the train's occupancy throughout its route.

## 2.6 The passenger's perspective

To see how all these elements work together, consider now the system from the perspective of a passenger. After consulting the system's maps, the passenger arrives at an origin station and confirms from the information outside that the station indeed serves his/her destination. The passenger then enters the station, and from the information described in Section 2.3 proceeds toward the correct platform. Upon arrival at this platform, the passenger consults the embarkation information described in Section 2.4 and proceeds to one of the gates serving her/his destination. Finally, upon the correct train's arrival, the passenger embarks through the open door at the gate and confirms with the interior sign next to said door that the door will indeed open for disembarkation at the destination of choice. Through the trip, the passenger may monitor the sign to anticipate the train's arrival at the intended station and correctly disembark.

Note that some of *PAXIS-XLT*'s features can be either omitted or simplified. Most existing and proposed passenger information systems are special cases.  Let us now turn our attention to train operations.

# 3. THE TRAIN OPERATION PROTOCOLS

The *XLT* operation protocols (*TOPS-XLT*) define how trains are configured and what each train does to pick up and deliver passengers at each station. Trains are assumed to stop at most once at each station, open and close their doors simultaneously, and then depart—as is conventional. The timetable is given. Three novel features of *TOPS-XLT* are:

a. Stopped *XLT*s can align any contiguous set of doors with a platform, provided they fit. Usually, *XLT*s will cover the complete platform and protrude beyond it at one or both of its ends. However, it is also possible for trains to cover only part of a platform (e.g., if the platform is very long), or not to cover it at all (e.g., if they skip the station).

b. *XLT*s can prevent disembarkation through some of the doors aligned with the platform by keeping them closed.

c. Embarkation for particular destinations can be restricted to arbitrary sets of gates. These sets can differ across stations, and at a given station also differ across successive trains.

The interactions between *XLT*s and station platforms are now described. To this end some *TOPS-XLT* terminology is needed. A unit is *aligned* with (or *visits*) a platform if all its doors face the platform and can be opened safely. An aligned unit is *disembarkation-ready* if its doors are opened and some are allowed for disembarkation.[3] An aligned unit is *presented for embarkation* to a particular destination if its doors are opened and the sign at the gate(s) facing the unit (and/or at the doors themselves) advertises embarkation for the destination in question.

In addition to the composition of each train in terms of units, a *TOPS-XLT* protocol must define the alignment, disembarkation and presentation decisions made for all units of all trains at all stations during the course of a day.

To simplify the *TOPS-XLT* decision-making process, we can artificially reduce the number of distinct units, stations and trains by treating them in groups of indistinguishable elements. In particular, consecutive units will be grouped into *sections* that will behave as units with doors operated in unison. Stations will also be grouped by type (e.g., as end-of-line, transfer, odd, even, etc.) so that a protocol's alignment, visitation and presentation decisions will be based on type alone. Finally, trains will be similarly classified by their unit/section composition and station types visited. Because these groupings can be as finely disaggregated as desired, no generality is lost by describing the protocols in terms of groups, as we shall do from now on. Finally, it will also be convenient to refer to the part of a station platform that is presented for a given destination type as a *segment* (for the particular destination type)—this is where passengers bound for destinations of the type in question would wait.[4]

Protocols can range from the simple to the very complex. At the simple end of the spectrum are *full-availability* protocols; i.e., where every aligned unit is made fully available for both, disembarkation, and for embarkation to all the destination types it visits. In these cases, the *destination labels* affixed to each unit (or section) are a function of the alignment decisions, and invariant across stations. For this reason the protocols will also be called *static*. For these types of protocols trains can be partitioned into *parts* composed of sections/units with the same destination labels.

---

[3] Usually all the open doors are available for disembarkation. In some cases, however, e.g., when a unit faces a different platform on each of its sides, some doors may be used for embarkation and others for disembarkation.
[4] The word segment is used because said part is usually continuous. Note that the segments of different station types will overlap when a train's sections are presented for more than one destination.

At the other end of the spectrum are *dynamic* or *partial-availability* protocols where disembarkation is full but presentation at some stations is partial. There seems to be no real reason to implement partial disembarkation policies, but restricting embarkations by hiding destination labels allows an agency to reserve space in a section for passengers boarding at downstream stations. These dynamic protocols can be *pre-planned* or *adaptive*. The latter require sophisticated sensing, control and passenger information systems. Static protocols, however, still offer considerable benefits, and we consider them in detail in the sections that follow. Dynamic protocols will also be discussed.

For maximum generality, the remainder of this paper assumes that passengers cannot change units while onboard. The ideas then apply to all types of trains, with and without passageways between adjacent units. Section 8 discusses what happens if this restriction is relaxed.

## 4. THE F/R FAMILY

The F/R protocols use one train type and two station types, which are labeled "F" (for forward) and "R" (for rear) for reasons that will become apparent. The protocols lengthen trains without restricting the passenger travel options in any way; and can be applied to a single *XLT* regardless of what other trains do--even if the trains pass one another and belong to different lines. As a result, the protocols can be conveniently applied to any set of trains, lines and directions one chooses. This flexibility should appeal to operators. The F/R protocols should also appeal to passengers because the protocols do not reduce the level-of-service. Even if applied to all the trains on a line, passengers can still use every station and board every train, thereby experiencing the same walks, waits and onboard travel times as if the F/R strategy was not used. In typical situations, an F/R *XLT* could carry about 33% more people.

Two versions of the approach are described below: the first is static, and the second dynamic and better adapted to the demand.

### 4.1 The static-homogeneous F/R-H protocols [5]

In this simple version, trains have four equal sections. They are statically labeled from front to rear with the destination types they serve: "F", "FR" or "RF", and "R". The two sections in the middle are in the same part of the *XLT* since both present F and R. Therefore, there are three parts to a train.

Figure 1 illustrates the protocol's decisions. It shows the *XLT*'s alignment at the two station types, and the destinations it presents. Note in particular how: (i) the shaded middle part of the train, which serves both destination types, visits every station; and (ii) the two end parts, which serve only one station type, are only aligned at stations of the corresponding type.

---

[5] This is an improved version of Avedisian (2017).

The figure also illustrates the passenger information system. Invariant segments on each platform indicate where passengers should wait for each destination type. Passengers going to a destination station of the same type as the embarkation station are allowed to board anywhere since all the train sections aligned at the embarkation station also visit all matching destinations. However, passengers traveling to opposite type stations need to embark through the segments of the platform that face the train's middle part because only this part will visit their destination type. These segments are shaded and labeled by destination type in the figure. It should thus be clear that the protocol allows all passengers to disembark where they want.

The protocol is fully defined once each station is assigned a type. This can be done arbitrarily without restriction. This flexibility allows us to balance the demands; e.g., by alternating designations if the demand is spatially uniform. The flexibility is also useful because by labeling the first station of a line "R" and the last "F", we ensure that trains protrude beyond the platforms of these stations on the side where tracks already exist. No expansion of end-of-line (EOL) stations is necessary. This is also true if service is bidirectional but the direction of travel can be reversed by operating the train backwards because stations can be renamed differently for each direction. All that is necessary is changing the designation turnaround station from "F" to "R" as the reversal takes place; and of course, changing the designations of the remaining stations to whatever works.

### How the F/R-H protocol increases the number of passengers carried

Let us define "capacity" as the maximum number of passengers that can flow through a subway line's maximum load point (MLP). Capacity is reached if during some time interval the *XLT*s are separated by minimum headways and all their cars register 100% occupancy at the MLP. Clearly, full occupancy is a worthy goal.[6]

The goal can be approximately reached by the proposed protocol. Unfortunately, the extent to which it is attained depends on the number of passengers with origins and destinations of the same type; and also on what they do, which we can surmise but the F/R-H protocol does not control. Although these passengers can board on any part of the train we expect them to prefer an end-section because these sections open and close their doors less frequently, offering a more serene environment. Thus, it is reasonable to expect that most passengers of this type will board an end-section unless the middle sections are anticipated to be less crowded. If this occurs, some of these passengers will gravitate toward the middle-sections and even out the car loads. This behavior is feasible because frequent commuters can anticipate from prior experience the expected loading of particular cars, and therefore queue up for them on the platform.

---

[6] There could be other goals; e.g., the maximization of total passenger embarkations or total passenger-kilometers from all O-Ds along the line, subject to no-overcrowding constraints. See Section 6.3 for an example.

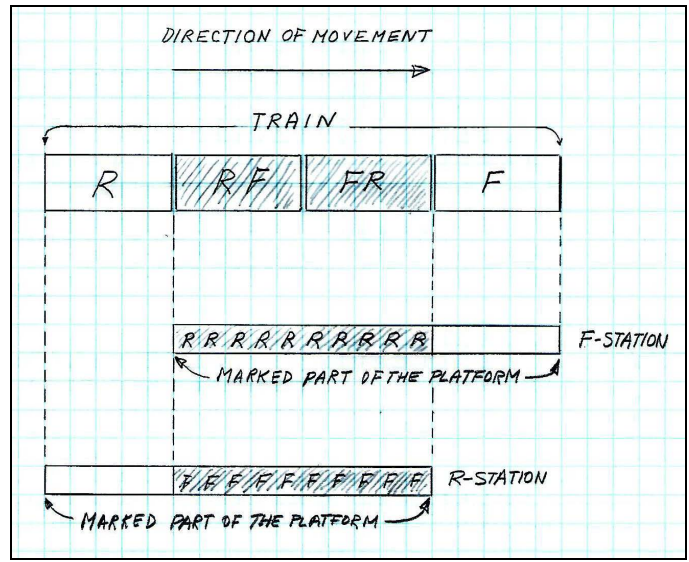

Figure 1. A train and two station types under the F/R protocol

With this passenger behavior trains will fill all the cars at the busiest link if two things happen: (i) there is enough demand of the matching type (F to F or R to R) to fill the two end sections of each train by the time the train reaches the busiest link; and (ii) there is enough demand of the other type to top off the middle part. These conditions should be usually met by any overcrowded subway during part of the rush so we see that the protocol can fill the four sections of the train. With a conventional protocol, however, trains can only have three sections, so only 3-full sections can be filled per train. The F/R-H protocol can therefore increase capacity by about 33% if the minimum safe headways of *XLT*s and ordinary trains are roughly equal, as is reasonable to expect.[7]

Of course, this potential may not be realized if there is not enough demand of the matching type to fill the end sections; or else if the demand distribution is unfavorable in other ways. Fortunately, this drawback can be mitigated as explained in the next subsection. The subsection will also show how capacity can be increased beyond 33% if the passenger distribution is favorable

## 4.2 The dynamic-inhomogeneous F/R-I protocols

These protocols include two improvements. First, the destinations are partially presented in such a way that every passenger is directed to a single section of the train, eliminating passenger choice. Second, the four train sections are allowed to have different lengths, which are then matched to the projected section loads.

---

[7] See Section 8 for some discussion on minimum headways.

The partial presentation scheme is as follows. The second section from the front, which visits stations of both types, only presents destinations "R", and does so only at F-stations. So this section only carries passengers going from F to R. Similarly, the third section only presents destinations "F" and only at R-stations, so this section only carries passengers from R to F. The remaining passengers, going from F to F and R to R, are forced onto the end sections. Thus, passengers have no choice.[8] This pattern induces a 1:1 correspondence between O-D pairs and sections. This is illustrated in Figure 2a by the markings inside the *XLT* boxes of each section. As a result, the passenger loads in each section can be approximately predicted from the O-D table. Also note from the figure how the markings on the platforms, which display the destination labels for each gate, present only one destination per gate {F, R or X} —the "X" meaning that the gate is not available for embarkation. This perfectly segregates passengers by destination on the station platform, unlike it occurred with the F/R-H protocol.

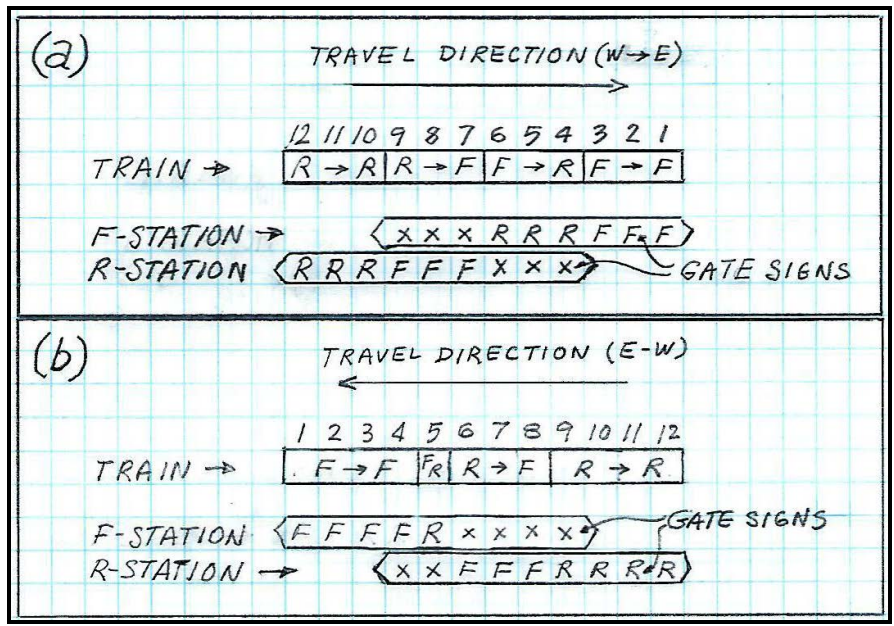

Figure 2. Gate signs displayed at the platforms of F-stations and R-stations in the two travel directions of a bidirectional subway line operated with the F/R-I protocol: an "X" indicates that the gates on that part of the platform are only allowed for disembarking.

The second improvement, flexible section lengths, allows us to set these lengths in proportion to the projected passenger loads. This, ensures an even passenger density across all sections, and guarantees that if one section is full all are full. No space is wasted.[9] Moreover, for favorable demand distributions, the two improvements combined can increase the line capacity well above the maximum possible with the F/R-H protocol. This is illustrated by the following example.

---

[8] Passengers do not have an incentive to disobey the signs because if they do they will likely find themselves in a more crowded car. And the critically important "no embarkation" signs can be physically enforced with barriers.

Consider an East-West bidirectional subway line that passes through a busy city center. The line is operated with the F/R protocol in both directions but even so, it cannot keep up with demand in either direction. Therefore, the subway agency wants to ensure that its trains are fully loaded when they flow through the two MLPs. The agency uses 12-car *XLT*s on platforms that can accommodate only nine, as in Figure 2a. Cars can operate their doors independently, so they are the system's units. They are numbered consecutively from front to rear as shown in the figure.

Table 1 below specifies the demand. It gives for each direction, the breakdown of passengers by O-D type observed at the MLP. Note how this distribution is uniform in the E to W direction but not so from W to E.

Table 1: Fractional distribution of passenger O-D types observed at the two MLPs of a bidirectional subway line. In parenthesis are the number of cars allocated to the sections serving each specific passenger type.

| Passenger classification by O/D type | | Fraction of passengers (Number of cars per section) | |
|---|---|---|---|
| Origin type | Destination type | E to W | W to E |
| F | F | 1/4  (3) | 1/3   (4) |
| R | R | 1/4  (3) | 1/3   (4) |
| F | R | 1/4  (3) | 1/12  (1) |
| R | F | 1/4  (3) | 1/4   (3) |

Since the assumed goal is to maximize capacity, an even passenger density at the two MLPs is a reasonable objective. To achieve it, the section lengths for each direction should be defined proportionately to the fractions of Table 1 for the corresponding direction. This is easy to do. The table shows that the E-to-W demand (and therefore the projected section loads) is evenly distributed amongst the four O-D's. Thus, for this direction, the four sections should be of equal length, with the train's middle part spanning one half of the train. Since trains have 12 cars, each section should have three cars. This is indicated by the parenthetical numbers on the

---

[9] The same guarantee cannot be offered by protocols such as the F/R-H that give passengers more leeway when choosing the section in which they will travel.

table's third column and by the diagram of Figure 2a. The same logic reveals that for the other direction the sections should have 4, 4, 1 and 3 cars, as indicated both, by the table's last column and by Figure 2b. The train's middle part now equals in length the other two parts.

Let us now see how the capacity improvement produced by the F/R-I protocol can exceed by a considerable amount the 33% maximum of the F/R-H protocol for favorable demand distributions. To do this, let us continue with the same example but now imagine that the total demand level is so low in the W to E direction that only the E to W direction requires an *XLT* protocol. We shall show that capacity then improves by 50%.

To see this, note from the diagram in part (b) of the figure that if the F-station rectangle representing the platform was shortened from the right by one unit, the platform would still accommodate the three forward sections of the trains, as required for embarkation and disembarkation by the F/R-I visitation plan. Therefore, the F-station platforms only need to be 8-cars long. Now note that the same happens but in reverse for the R-stations. Their platforms too, only need to be 8-cars long. If platforms had been this short to begin with, ordinary trains could only have 8 cars but our 12-car *XLT* trains could continue to operate and still be filled. Thus, for our demand distribution, the FR-I protocol would increase the capacity of an ordinary system by 50%.

This example shows that the demand distribution by O-D type can impact performance considerably. Now, since said distribution can be affected by the allocation of types to stations, it stands to reason that such allocation should be optimized systematically. In view of this, Section 6 will introduce a *TOPS-XLT* optimization framework that will allow this kind of analysis. Before this is done, however, Section 5 introduces a protocol family that uses even longer *XLT*s by slightly relaxing the level-of-service standards.

## 5. THE F/T/R FAMILY

The TOPS-XLT family about to be presented can improve capacity by 100%. This is achieved by using trains that are twice as long as the platforms, but at the cost of slightly limiting the passengers' travel options. The protocol does not restrict travel to/from important stations such as transfer stations, but it precludes direct travel between 50% of the remaining station pairs. Affected passengers can either transfer or change the station at one end of their trips. The latter increases their access distance. For the sake of brevity only the static homogeneous version of the family is described, i.e. where *XLT* sections are identical, every door opens when aligned and destinations are fully presented.

Like the F/R protocols, the F/T/R protocols can be applied to individual train runs without coordination with other trains. Although *XLT*s continue to have four sections, they now have four parts instead of three. Another difference is that a third station type "T" is introduced for stations deemed to be so important that they must be directly connected in both directions with every other station. The label "T" is used because transfer stations should logically be in this class. Thus, the station types are now labeled "F", "R" and "T". The main advantage of all these changes is that the station platforms then only need to span two *XLT* sections—not three.

Figure 3 below shows how the protocol works using the conventions of Figure 1. Sections have been labeled with the station types they visit—and present. Note how each section is a separate part. Also note from the shown alignments that the F-stations are only visited by the first two sections of each train, the T-stations by the middle two, and the R-stations by the last two. As a result, passengers embarking at F-stations can only go directly to F-stations or T-stations—although they could reach R-stations by transferring at a T-station. Similarly, passengers embarking at R-stations can only go directly to R-stations or T-stations. However, passengers embarking at T-stations can go anywhere.

The passenger information system should tell passengers outside the station which station types are served directly from the station, and which require transfers. Then, to ensure passengers reach their desired destinations, *PAXIS-XLT* must direct them in the station to the proper train sections. Obviously, passengers going to a station of the same type can embark anywhere along the platform, and do not require specific directions, since the two sections presented to them visit every station of the same type.

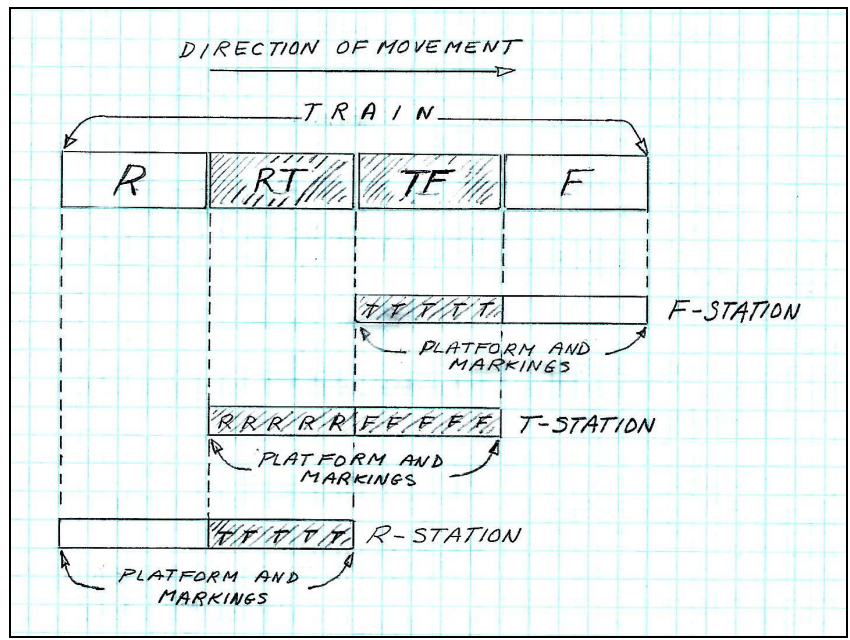

Figure 3. A train and three station types under the F/T/R protocol. The diagram shows the markings on the platforms and how a stopped train aligns itself with the platforms of the three types. The train is assumed to move from left to right.

The remaining passengers must embark only through segments facing train sections that will visit the desired (non-matching) destination, as shown in Figure 3. At F- and R-stations "T" is the only non-matching destination type on offer. Figure 3 shows that if passengers going to "T" from F- and R-stations embark where the gates are labeled "T", they will board a section that visits their destination. At T-stations the non-matching destinations are "F" and "R". Figure 3 shows how in this case too, if each non-matching passenger embarks through a gate labeled with his/her destination, (s)he will be able to disembark at the destination of choice. Thus, the protocol works as intended.

<u>How the F/T/R protocol increases the number of passengers carried, potentially doubling it</u>

As in Section 4.1, we reasonably assume that commuters with a choice of sections will board those anticipated to be most comfortable. In particular, we surmise that when facing such a choice, commuters going from F to F (or R to R) should favor the more tranquil end-sections.

As with the F/R protocol, all cars are then filled by the MLP if: (i) the traveler demand from F to F (and R to R) can fill the train's end-sections; and (ii) the overflow of this demand, combined with the rest, can fill the middle sections. The F/T/R protocol would then send through the MLP a full four-section-*XLT*. By contrast, an ordinary train could only have two sections. Therefore the F/T/R protocol can double train occupancy and the system's capacity.

Of course, the larger improvement in capacity compared with the F/R protocol requires some passengers to either transfer or access/egress the system at a second-best station. To quantify this inconvenience consider a line with homogeneous demand where the stations are evenly spaced and consecutively labeled {…F, R, T, F, R, T…}. Then, if no passenger chooses to transfer, the excess access and egress distance/time turns out to be less than 4% of the total.[10] However, the actual increase in passenger travel time should be even smaller for two reasons. First, because for some passengers transferring may be quicker than adding to their walks; and second, because station labels can be assigned intelligently to increase the number of matching-destination passengers. In a real decision-making situation the extra passenger cost, together with the extra cost for the additional rolling stock required by the *XLT*-trains, would have to be weighed against the benefit of the protocol's increased capacity.

As with the F/R protocol, three improvements can be made: (i) inhomogeneous sections can be defined for use in both mono- and bi-directional lines; (ii) destinations can be partially presented to eliminate passenger choice and better control train loads; and (iii) stations can be labeled intelligently, to maximize capacity and level-of service while still allowing trains to turn around seamlessly at EOL stations. The next section explains how these improvements can be systematically selected.

---

[10] Affected passengers are about 2/9 of the total because only 2 of the 9 possible O-D pairs are not served directly. Furthermore, affected passengers can choose the second-best station to be either at the origin or destination end of the trip; i.e., where the extra distance is less. With this in mind, simple geometric probability calculations reveal that if the station spacing is uniform, then the average extra access distance endured by affected passengers is only 1/6 of the spacing. Moreover, since this extra travel applies only to 2/9 of the population, the average extra distance across all passengers is (1/6)(2/9) = 1/27 < 4%.

## 6. PROTOCOL SPECIFICATION AND OPTIMIZATION

This section will present a framework to specify and optimize TOPS-XLT protocols. The goal is not to present the best framework, but to introduce a proof-of-concept idea to inspire further thinking. The development of improved analysis methods could be a research thrust in the field of subway scheduling.

In the proposed framework, the protocols' decisions will be identified with tables of zeroes and ones, or equivalently with sets of binary decision variables, which are subject to feasibility constraints. Section 6.1 defines the variables and Section 6.2 the constraints. To perform an optimization, an objective function also needs to be formulated. This is complicated because an arbitrary set of protocols for the trains dispatched in a day produces a level of service that depends on both the demand and the specifics of the protocols in complex ways—involving a time-dependent assignment of passengers to trains and sections, while considering transfers and section occupancy constraints. Fortunately, if we confine our optimization to within a family of protocols that provide similar passenger connectivity, the in-system travel time component of the level of service is the same across family members and can be dropped from the formulation. Section 6.3 illustrates this idea. Finally, Section 6.4, suggests a way to address the general optimization problem.

### 6.1 Definitions

To start let us define individual ID's $\{t\}$ for all trains operated. These ID's can be numerical or categorical. Next, we label stations $s$ and the train-units $m$ in each train by consecutive integers that increase in the direction of travel for stations, $\{s = 1, 2\ldots S\}$, and from front to rear for trains, $\{m = 1, 2\ldots\}$. We also define $C$ station types $\{i\}$ and $K$ train types $\{k\}$, both of which can be alphanumeric. The label $j$ will be reserved for station types as well. The units of each train type are identified by the double subscript $km$, so their common length and passenger holding capacity are denoted: $l_{km}$ and $c_{km}$, respectively. The number of units of type-$k$ trains is denoted $M_k$. Finally, we define $d_i$ as the length of the shortest platform of type-$i$.

With these elements defined, any *TOPS-XLT* protocol can now be specified with a set of tables containing binary decision variables (b-variables). To do this, first assign a type to each station and train with the following two classification tables.

1. <u>Station classification table</u>: b-variables $\delta_{si}$ that indicate whether station $s$ is of type $i$. Since every station is of one type, the variables must satisfy $\sum_i \delta_{si} = 1$.

2. <u>Train classification table</u>: b-variables $\varepsilon_{tk}$ that indicate whether train $t$ is of type $k$. Since every train is of one type, the variables must satisfy $\sum_k \varepsilon_{tk} = 1$.

Next, decide how each train type is partitioned into sections of consecutive units. The number of sections is denoted $N_k$. The sections are identified by a numerical index $n = 1, 2 \dots N_k$ that increases from front to rear. The corresponding table is:

3. <u>Section definition table</u>: b-variables $u_{kmn}$ that indicate if unit $km$ is in the $n^{\text{th}}$ section.

To ensure that all sections are composed of consecutive units the b-variables for any unit pair $(b, b')$ such that $1 \leq b < b' \leq M_k$ are required to satisfy:

$$u_{kbn} \times u_{kb'n} = \prod_{b \leq m \leq b'} \{u_{kmn}\} \qquad \text{(sections are composed of consecutive units).}$$

Finally, we must decide what trains do at each station type. Four decisions need to be made: stop/skip; alignment; visitation; and presentation. The corresponding tables are:

4. <u>Stop/skip table</u>: b-variables $s_{ki}$ indicate if type-$k$ trains stop or skip stations of type $i$.

5. <u>Alignment table</u>: b-variables $a_{kni}$ indicate if section $n$ of type-$k$ trains aligns at stations of type $i$.

6. <u>Disembarkation table</u>: b-variables $v_{kni}$ indicate if section $n$ of type-$k$ trains opens the doors at stations of type $i$.

7. <u>Presentation table</u>: b-variables $p_{knij}$ indicate if section $n$ of type-$k$ trains presents type-$j$ destinations at stations of type $i$.

The above tables completely describe a protocol. Additional tables do not have to be introduced for the passenger side because the *PAXIS-XLT* signs on the platform gates follow directly from the trains' alignment and presentation decisions.

## 6.2 Feasibility considerations

For a *TOPS-XLT* protocol to be feasible it must satisfy some feasibility constraints. They are:[11]

---

[11] These are not meant to be the final word. Additional inequalities can be introduced if a situation demands it; e.g., if one wishes to designate doors on one side of the train for embarkation only and doors on the other side for disembarkation.

$$u_{kan} \cdot u_{kbn} = \prod_{b \leq m \leq b'} \{u_{kmn}\} \qquad \text{(sections are composed of consecutive units)} \quad (1)$$

$$s_{ki} \geq a_{kni} \qquad \text{(sections are aligned only if the train stops)} \quad (2)$$

$$\prod_{b \leq n \leq b'} \{a_{kni}\} = a_{kai} \cdot a_{kbi} \qquad \text{(aligned sections are consecutive)} \quad (3)$$

$$\sum_{n,m} \{a_{kni} \cdot u_{kmn} \cdot l_{km}\} \leq d_i \qquad \text{(aligned sections fit along the platform)} \quad (4)$$

$$a_{kni} \geq v_{kni} \qquad \text{(section doors opened only if section is aligned)} \quad (5)$$

$$p_{knij} \leq v_{kni} \cdot v_{knj} \qquad \text{(destination presented only if the section opens its doors} \quad (6)$$
$$\text{at the current and destination stations)}$$

Depending on the application, some of the variables in our tables may be given as data, and the rest left to be optimized subject to possibly additional constraints specific to the situation. For example, for the homogeneous-static protocols of sections 3 and 4, $N_k$ is fixed at 4 and Tables 3-7 are given data. The only choices in these cases are in Tables 1 and 2. For these we may introduce EOL constraints such as $\delta_{si} = 1$ if $\{(s = 1$ and $i = R)$ or $(s = S$ and $i = F)\}$.

For dynamic-inhomogeneous *TOPS-XLT* protocols such as the one in Section 4.2, Tables 3 (section sizes) and 7 (partial presentation scheme) would also be treated as decision sets. In this case, we might want to introduce constraints to guarantee minimum presentation standards; e.g., specify that each destination is presented at each station by at least one section $\{\sum_n p_{knij} \geq 1\}$ or, as was proposed in Section 4.2, by exactly one section $\{\sum_n p_{knij} = 1\}$.

Maximum occupancy constraints based on demand O-D tables can also be introduced if one wishes to operate the system with a good level of service and comfort. The only difference from conventional analysis is that sections should be treated now as if they were separate vehicles. To show how this can be done an example is given below.

## 6.3 Example: Configuring and metering an oversaturated subway line

We explore here how best to organize a subway line that is operated with regular headways of duration $H$ [h], using trains composed of identical units with maximum occupancy $c$ and length $l$. Because the system is severely oversaturated the operating agency plans, both, to adopt the F/R-I protocol and regulate passenger access with optimally calibrated meters at stations. The agency's wishes to configure the protocol and set the metering rates so trains can operate without overcrowding while serving as many people as possible. For brevity, we formulate this problem assuming a steady state but the ideas can be easily generalized.

Given is a steady-state O-D demand table $\{A_{ss'}\}$ [pax/h]. Thus, the demand rate at station $s$ is $A_s = \sum_{s'} A_{ss'}$ [pax/h]. We look for the rates $E_s \leq A_s$ at which the meters should allow people to enter each station. On account of social equity, the agency also introduces minimum entrance rates $M_s$ for each station, so the $E_s$ must satisfy:

$$M_s \leq E_s \leq A_s.$$

The objective is maximizing the total number of people served subject to the above constraint; i.e.:

$$max\{\textstyle\sum_s E_s\}$$

In addition to setting the metering rates, the agency can also configure the trains and classify the stations. Therefore we treat the sets $\{\delta_{si}\}$ and $\{u_{mn}\}$ of Tables 1 and 3 as decision variables. The remaining tables (which define the F/T-I protocol) are given. Of note, we are dropping the train-type subscript "$k$" and ignoring Table 2 because our protocol has only one train type.

To complete the formulation, we need constraints to prevent overcrowding; i.e., that no section overflows after leaving any station. To this end, the following auxiliary variables are defined:

$$C_n = c\textstyle\sum_m u_{mn} \qquad\qquad \text{(defines the sections' maximum occupancies)}$$

$$E_{ss'} = E_s A_{ss'} / A_s \qquad \text{(defines the passenger entry rates by destination, assuming FIFO order)}$$

In addition, it is convenient to define whether passengers going from $s$ to $s'$ use section $n$ or not. A b-variable $\Delta_{nss'}$ is introduced for this purpose. To understand how it will be defined, recall that the F/R-I protocol assignment is "all-or-nothing" so that the fraction of the flow going from $s$ to $s'$ that boards section $n$ is either zero or 1. Note as well that the origin-destination type corresponding to $s$-$s'$ is the only $i$-$j$ pair for which $\delta_{si}\delta_{s'j} = 1$. Thus, the fraction of passengers with this origin-destination type that board section $n$ is $p_{nij}$ and we can define our binary indicator as:

$$\Delta_{nss'} = \textstyle\sum_{ij} \{\delta_{si}\delta_{s'j}\, p_{nij}\} \text{ if } s' > s; \text{ or else } \Delta_{nss'} = 0 \qquad\qquad \text{(defines } \Delta_{nss'})$$

To prevent overcrowding, we stipulate that the passenger load in each train section $n$ after departing every station $s$ must be below or equal to the section's maximum occupancy. Since said load is the sum of embarkations at all previously visited stations $\{z: z \leq s\}$ of passengers with destinations $\{s': s' > s\}$, the load is: $\sum_{z \leq s} \sum_{s' > s} \{E_{zs'}\Delta_{nzs'}\}$. Thus, our overcrowding constraints are:

$$\textstyle\sum_{z \leq s} \sum_{s' > s} \{E_{zs'}\Delta_{nzs'}\} \leq C_n \qquad\qquad \text{(no overcrowding constraints)}$$

This concludes the formulation. It is a mathematical program that can be solved with either standard or customized optimization methods.

## 6.4 The general optimization problem

In a general case, we may want to treat all the tables as variables. Although this is not easy, it may be attempted by making separate decisions at two levels. An upper level set that would include all train-related decisions and a lower level set with all station- and door-related decisions. The upper set defines the number of train types $K$, their composition $\{M_k, l_{km}, c_{km}\}$, their sequence (Table 2), and their time-space trajectories including the stopping and alignment particulars (Tables 4 and 5). The lower set defines the number of station types $C$, the station labels (Table 1), the trains' section lengths (Table 3), and their door opening and presentation decisions (Tables 6 and 7).

In section 6.3, the upper set was given, and only Tables 1 and 3 of the lower set were variables. Tables 6 and 7 of the lower set were considered fixed (as per the F/R-I rules) but they could have been allowed to vary. Of course, we can be much more general. Conceivably we could formulate a lower-level problem to maximize some objective (e.g., capacity) conditional on the upper-level variables; and then choose the latter by solving a bi-level optimization program for some appropriate objective function, with the lower-level problem as a constraint. Needless to say, this is easier said than done. Fortunately, there are some cases when the complete problem simplifies considerably, and this is explained in the next section.

# 7. SIMPLIFIED MODELING

This section shows how to create and optimize protocols for systems that use a single unit type as this allows for considerable simplifications. As a result of these simplifications, static protocols can be specified with just a few decision variables and visualized with a simple diagram (Section 7.1). The results can be refined with dynamic presentations. Thus, better protocols can be found both for single trains (Section 7.2) and combinations (Section 7.3).

## 7.1 Modeling and optimizing a single train

In what follows, and without loss of generality, each unit is its own section. The protocol specification task is considerably simpler than in the general case for three reasons: (i) With identical units and $K = 1$, train composition is specified by a single variable $M$ and the subscript $k$ can be dropped; (ii) Table 3 is fixed, since $u_{mn} = 1$ if $n = m$ or zero otherwise; and (iii) Tables 6 and 7 are functions of Table 5, since $v_{ni} = a_{ni}$ and $p_{nij} = a_{ni} \cdot a_{ni}$. Therefore, aside from $M$, the only decision variables are Tables 1, 4 and 5—Table 2 plays no role for a single train. Apart from Table 1, which deals with stations, the remaining train-related information can be encapsulated in a stacked-bar chart diagram depicting the $XLT$ alignments.

The top bar of such a chart shall represent the train and be of length $M$. To capture the information in Tables 4 and 5 the diagram shall also include one bar of length $d_i$ for each station type. As in Figures 2 and 3, each of these bars should be positioned so it spans the units of the train that visit the corresponding stations. Therefore, the bars neatly identify the set $\{a_{ni}\}$; i.e., all the information in Table 5. If in addition, we use the convention that a zero-overlap bar signifies skipping the station, then the diagram can also encapsulate Table 4. Figure 4 shows an example with four station types.

The chart can also summarize the presentation scheme (Table 7). Recall a unit is presented for embarkation to all the destinations types it visits; i.e., all station types whose bars intersect a vertical line emanating from the unit. Therefore, a train part (which presents a fixed set of destination labels) is any interval on the top bar whose vertical lines intersect bars of the same type; and the part's destination labels are those of the intersected bars. All this can be seen at a glance from the chart, as shown by Figure 4 which summarizes the destination labeling information across the top of the train.

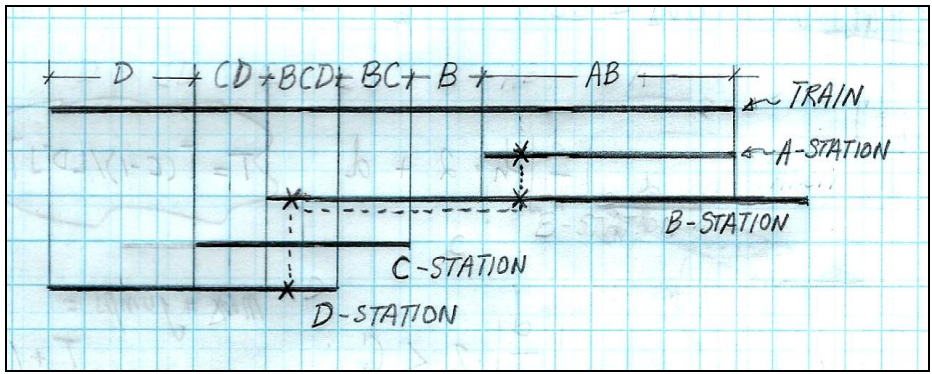

Figure 4. Alignment diagram for a static protocol with $C = 4$ station types and a six-part-train with $M = 19$ units. The dotted lines depict the path of a passenger trip between stations of types A and D. The displacements of each bar relative to the train are, from bottom to top: $b_i = \{8, 10, 21, \text{and } 19\}$.

The chart also describes the routes passengers with different O-D's can take to complete a trip. Any such route is described by a step-wise path on the diagram. Figure 4 illustrates the idea with a trip from "A" to "D". The path starts with a vertical dotted segment between A and B denoting the passenger's first leg of the route from A to B in part AB of the train. It continues with a horizontal dotted segment entirely within the B-bar, denoting the passenger's walk along the B-platform to effect the transfer. And it concludes with another vertical dotted segment between B and D, denoting the final leg of the passenger's route from B to D in part BCD of the train. Note that our path is directionless so it also describes a feasible travel route from "D" to "A". It should be clear that our directionless paths are feasible if and only if they are composed of horizontal steps that stay confined within a single bar and vertical steps that start and end at a bar—even if they intersect other bars. (The exception is paths between stations of the same type, which consist of a single dot on the corresponding bar.) A goal when choosing a route for an O-D pair is to find the path with the fewest transfers. The reader can verify by inspection of Figure 4 that O-D pairs A-C and A-D require one transfer and the rest require none. Finally, note that any instance of an alignment diagram is characterized by $M$ and the horizontal displacement of each bar relative to the train, measured by the number of units, $b_i$, separating the front of the bar and the back of the train. The caption of Figure 4 shows the values of these variables. For an aligned train they can take any integer value from $b_i = 1$ to $b_i = M+d_i-1$. (The value "0" can be used for skipped station types.) This characterization shows that a single-train static protocol is completely defined by $M$ and the finite set of bounded integers $\{b_i\}$.

If we use these variables (and Table 1) instead of the more numerous variables in Section 6.1, an exhaustive search becomes possible. This can be done in two nested steps. First, optimize (or exhaustively search for) $M$ and the $\{b_i\}$ conditional Table 1;[12] and then perform an outer search to optimize Table 1.[13]

Dynamic refinements to better distribute the resulting passenger loads along the train can also be found in a simpler way. Note, once the bars are chosen our $XLT$ is divided into parts (labeled $\pi$) that serve common destination sets; see Figure 4. Then, if we (reasonably) require all the units in each part to present the same set of destinations at each station, we can specify the partial presentation scheme in terms of these (few) parts; e.g., with b-variables $\{p_{\pi ij}\}$ instead of the many variables in Table 7. In the author's experience, greedy algorithms that determine $p_{\pi ij}$ one O-D pair at a time can notably improve the passenger distribution.

The next section presents a simple family of static protocols for applications where one wishes to use a common (minimum) platform length for all the stations. The family is fully specified with only two parameters, is easy to optimize and improves on the basic protocols already presented.

## 7.2 The S-protocols for a single train

Let the common platform length be $d$ [units]. Consideration that if a protocol for this type of problem is to prevent any passenger to require more than $T$ transfers, then the train length $M$ [units] must satisfy:

$$M < (2+T)\,d. \tag{7}$$

---

[12] Conditioning on Table 1 is necessary so the platform lengths $\{d_i\}$ can be defined

To see this note that the bar-diagram of a feasible protocol must have a passenger path with $T$ transfers or less that connects the leftmost and rightmost bars of the diagram. Furthermore, the bars on this path must overlap with their neighbors by at least 1 unit, so travel is possible for each of the trip's legs. Therefore, the horizontal separation between the left end of the leftmost bar and the right end of the rightmost bars (i.e., the useful length of a train, $M$)[14] must be strictly less than the sum of the lengths of the involved bars, which is $(2+T)d$. Hence, (7) holds.

Note that the train lengths of the F/R and F/T/R protocols fall well short of the right side of (7). In view of this, a family of protocols that alleviates this shortcoming is now presented. The family includes the protocols of earlier sections as special cases.

The family's defining property is that the series of $\{b_i\}$ increases uniformly in steps of $h$ units from $b_i = d$ at the bottom of the chart to $b_i = M$ at the top. (For this reason the family is called the "Step-family" or "S-family" for short.) Clearly, if the step size $h$ and the number of classes $C$ are given we can draw the chart. Thus, these two parameters define the protocol. It will be convenient to express $h$ in units of $d$ and replace it by its inverse i.e., by the rational ratio $d/h \equiv D > 0$. Think of $D$ as the number of steps in a platform; or as the length of the platform measured in units of $h$. We shall refer to individual protocols in this family by the notation "$S(C, D)$".

The family is so simple that formulas for train length and for the required number of transfers for the worst-case trip can be obtained from the chart's geometry. Regarding train length, consideration shows that the length of an $S(C, D)$ train measured in platform lengths, $M/d$, is:

$$(M/d) = 1 + (C-1)/D \qquad \text{(train length).} \quad (8)$$

Now consider the required number of transfers for a worst-case trip. To find a formula, first note from the geometry of the bar chart that the maximum number of station types that can be connected with $T$ transfers or less is:[15]

$$C_T = 1 + (T+1)\lfloor D^- \rfloor \qquad \text{(maximum number of $T$-connected classes).} \quad (9)$$

Now, solving for $T$ in (8) we find the following for the number of transfers of a worst case trip:

$$T_{worst} \geq (C-1)/\lfloor D^- \rfloor - 1 \qquad \text{(transfers for worst case trip).} \quad (10)$$

---

[13] This search can be eliminated at the cost of complexity by assigning a different station type to each station.
[14] Of course, trains could be longer, but the extra units would be useless as they would never be aligned.
[15] The notation $\lfloor D^- \rfloor$ stands for the largest integer *strictly* smaller than $D$; i.e. $\lfloor D^- \rfloor \equiv \sup\{x: x \in \mathbb{N}, x < D\}$

By entering with (9) in (8) we also find an expression for the maximum length of *XLT* trains that can provide connectivity with *T* transfers or less:

$$(M/d)_T = 1 + (T+1) \lfloor D^- \rfloor / D \qquad \text{(maximum train length with } T \text{ transfers).}  (11)$$

Note, by choosing *D* to be a large number, (11) can approach (7) without actually reaching it.

Let us now analyze some examples. Figure 5 displays bar charts for the $S(2, 2)$, $S(3, 2)$, S(3, 3) and $S(7, 4)$ protocols. Note, the first of these is simply an F/R-I protocol with three equal train parts, such as the one in Figure 2b, and the second the F/T/R protocol. The dotted lines on S(3, 2) and S(7, 4) depict a feasible route for two passengers requiring a transfer.

Note from the figure how $S(2, 2)$ increases train length by 50%, with zero transfers, and how S(3, 2) increases train length by 100% with one transfer for the worst case trip. Both results agree with (7) and (9) and with our earlier discussion of the F/R-I and F/T/R protocols. More importantly, however, note how the third and fourth protocols respectively increase train length by 66% (with zero transfers) and by 150% (with one transfer). As such these protocols outperform the F/R-I and F/T/R protocols if the demand is favorable. This is not to suggest that they are the best. They are not. Optimality depends on the demand and a chosen objective function.

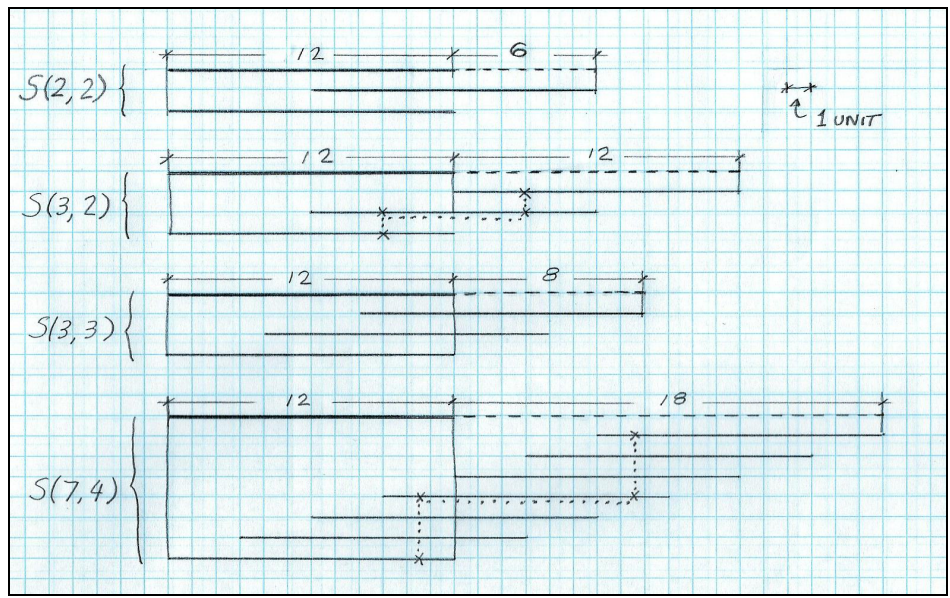

Figure 5. Four examples of S-protocols. The two numbers along the top of each train-bar are the number of units that can be lined up along a platform (solid line) and the extra units allowed by the protocol (dashed line). The dotted stepwise lines are paths taken by worst-case transferring passengers.

**7.3 Protocols with multiple train types**

Multiple train types can be used to further increase train lengths while improving connectivity and train speed. Both effects are discussed below.

<u>7.3.1 Increasing train length and connectivity</u>

We show here how train lengths can be increased beyond the bound (7) by using multiple trains. A first example will show that *XLT*s can provide perfect connectivity ($T = 0$) and be twice as long as ordinary trains A second example will show that with $T = 1$, *XLT*s can be three times as long. Both multi-train protocols improve on the protocols described in Sections 4, 5 and 7.2.

<u>The S(3, 2)/3 (a.k.a. F/T/R/3) protocol</u>. The first example is a three-train-type version of S(3, 2); i.e., of F/T/R. The new protocol improves on F/T/R because it provides total connectivity without transfers. The protocol uses three station types (F, T, R}, five station subtypes (A, B, C, D, T} and three train types (1, 2, 3}. The latter are dispatched with a regular rotation {1, 2, 3, 1, 2, 3…}.

Figure 6 displays the F/T/R/3 bar chart. As in previous figures the top bar, which spans 8 units, is the train; and the bars below, which span 4 units each, depict the station alignments. Clearly, the *XLT*s can be twice as long as ordinary trains. Because there are three train types, a separate set of bars is used for each type. Note the three sets of bars are identical. The only difference across train types is in the station subtypes used to define the station classes F and R: {Type 1: F=(A,B), R=(C,D)}; {Type 2: F=(A,C), R=(B,D)}; {Type 3: F=(A,D), R=(B,C)}.

The new protocol improves on S(3, 2) and F/T/R because the figure's trains are just as long; and because, as the reader can verify, passengers of every O-D type can now travel directly if they wait a little. For example, travelers from D to B can avoid a transfer by waiting for a train of type 2. Thus, the new protocol eliminates the requirement for transfers while still doubling train length. Its main downside is that the tracks of the station at the beginning of the line would have to be lengthened somewhat--since for every possible subtype assigned to that station, at least one train type would have to protrude beyond the upstream end of the station's platform.

<u>The S(5, 2)/2 protocol</u>. The second example is a two-train-type version of the S(5, 2) protocol, which uses very long trains (3-times ordinary) but requires some passengers to make three transfers. With the proposed protocol the two train types are of equal length (still 3-times ordinary) and are dispatched in alternating fashion from the head of the line. The trains do not pass. Therefore, the trains arriving at every station alternate in type. We shall show below how this arrangement reduces the maximum number of transfers from three to one.

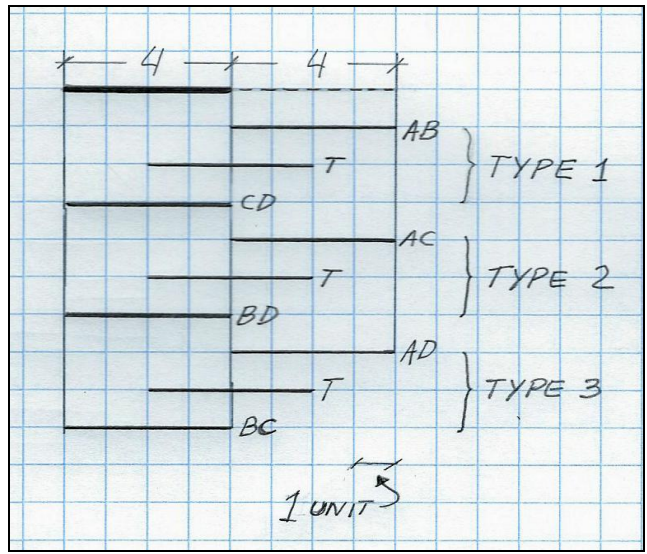

Figure 6. The F/T/R/3 protocol showing its three train types and station type definitions. {Type 1: F=(A,B), R=(C,D)}; {Type 2: F=(A,C), R=(B,D)}; {Type 3: F=(A,D), R=(B,C)}.

Figure 7 displays the S(5, 2)/2 bar chart. Note how trains are 12 units long and platforms 4 units long; thus, our *XLT*s are three times longer than ordinary trains. Both *XLT* types follow nearly identical S(5, 2) protocols. The only difference is a swap between bar-labels B and D.

This swap is key because it expands the set of destinations passengers can access when transferring. This occurs because by changing trains at B (or D) a passenger has access to any destination that overlaps with *either* of the two B-bars (or D-bars), and not just with the destinations that overlap with the bar of the train the passenger was riding. Because the two B-bars (D-bars) are staggered the set of accessible destinations is expanded.

One way of visualizing this is by including dotted connectors between bars of the same type and allowing these connectors to be part of the step paths describing the passenger route traveled. One such connector is displayed in Figure 7. These connectors are used when passengers change train type.

Consideration of the step paths (and connectors) for the ten possible O-Ds of our example reveals that every O-D pair can be reached with at most one transfer—instead of three. For example to go from A to E a passenger can take two possible paths: {A-B (train 1): B-E (train 2)} and {A-D (train 2): D-E (train 1)}. As a result of this choice set, the passenger in question can board the first train to come and then transfer to a train of opposite type.

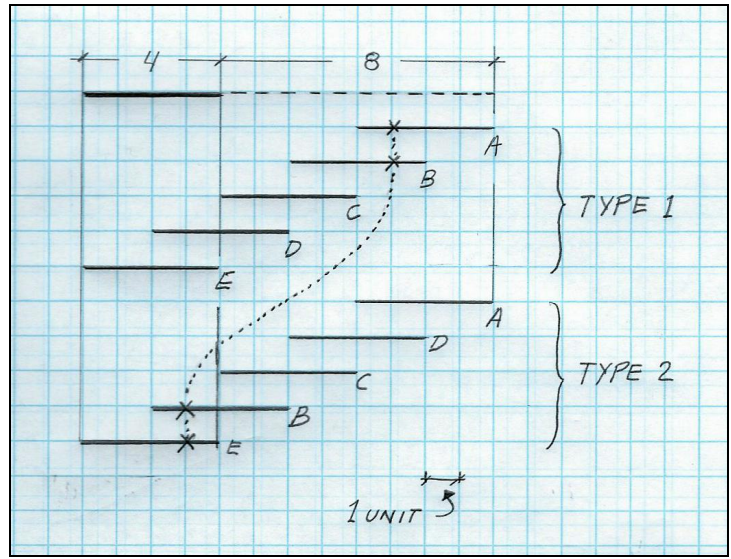

Figure 7. The S(5, 2)/2 protocol showing the station labels for its two train types. Labels B and D have been swapped across types. The dotted curve is the path of a passenger transferring across train types.

Finally note that by using label A for the top two bars of the alignment diagram, and label E for the bottom two, we can assign these labels to the EOL stations and prevent every train from protruding beyond the end of the EOL platforms. Therefore these stations do not have to be expanded to accommodate our *XLT*s. More importantly, the example has shown that train length can be tripled while allowing the worst-case passengers to complete their trips with only one transfer. This improves on the S(3, 2) and F/T/R protocols, which also use one transfer but can only double train length. It also improves on the one-transfer S-protocols, which cannot triple train length, as per (10).

### 7.3.2 Increasing train length and speed

Agencies often use trains of different kinds on long lines. This allows them to reduce the number of stops some trains make in their long journeys, speeding up passenger rides and improving level of service. A common example is the use of local and express trains on the same line. Another example is the odd-even strategy for serving alternate stations.

All this can also be done with *TOPS-XLT* protocols. Simply, associate one (or more) *TOPS-XLT* train type(s) with a distinct subset of stations served, specifying the protocols as if the stations outside the subset did not exist. Do this for as many station-subsets as desired.

To be concrete, consider a system with three station types {A, B, T} and two subsets {A, T} and {B, T}, where each subset is served by a unique train type using the zero-transfer S(2, 2) protocol. The bar chart on top of Figure 8 illustrates the setup. The letters along the bottom are the labels assigned to the 15 stations of the line--from left to right in the direction of travel. This setup speeds up trains because they only stop at 9 stations instead of 15. The EOLs do not have to be expanded if we choose the first station to be of type T and the last of type A or B, as illustrated. A downside is that the strategy is no longer zero-transfer like the single-train version, since passengers going from A to B and B to A need to transfer at the T-stations. This may not negate the advantage of shorter riding times, however, if the subway line and people's trips are long enough, as may occur in a megacity.

The chart on the bottom of Figure 8 uses the same conventions to illustrate how a skip-stop strategy can be implemented with the one-transfer S(3, 2) protocol. The line now has 29 stops, but trains only make 16. By allocating labels C or D to the first station in the string, and labels A or B to the last, the EOL stations can be kept as is. Curiously, in this case the maximum number of transfers does not increase compared with the single-train protocol.

The S-protocol family examined throughout Section 7 was homogeneous: it had even staggers and equal train parts. This can be somewhat limiting when the O-D table is irregular—even though the O-D table can be controlled by judiciously classifying the stations. The family was also static. These simplifying features can perhaps be relaxed with iterative searches that use a protocol in the S-family as an initial solution.

## 8. DISCUSSION

This paper has described a system (*OS-XLT*) for operating extra-long trains (*XLT*s) on subway lines. Two basic examples were introduced in Sections 4 and 5. Also discussed were more refined protocols, such as the F/R-I protocol of Sec. 4.2, which can even out the passenger load across all the train units so no cabin space is wasted. The paper also described in Section 6 an analysis framework to specify and numerically optimize the protocols. This framework was simplified in Section 7 for a common class of problems.

With *OS-XLT*, trains much longer than ordinary trains can be operated without changing the system's infrastructure. As an illustration, Section 7 showed how train length can be doubled without using transfers (Figure 6) and tripled if some passengers are asked to make one transfer (Figure 7). This section also showed how *XLT*s can be flexibly scheduled to provide any kind of service ordinary trains can perform--such as skip-stop service (Figure 8).

Because the paper implicitly used two mild assumptions, we now discuss their implications and then close with a suggestion for additional work.

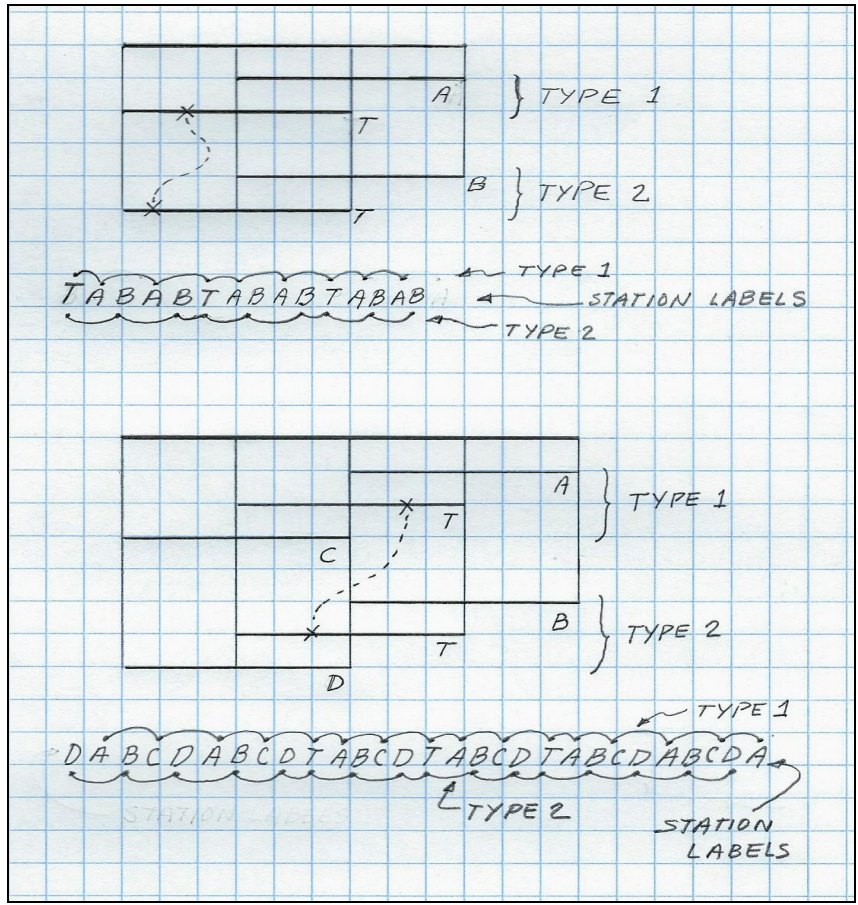

Figure 8. Two-train, skip-stop strategies using *TOPS-XLT* protocols: top, S(2, 2); bottom, S(3, 2). The string of labels below each chart is the station classification (Table 1). Trains travel from left to right. The curves above and below each string depict the stops of each train type.

The first assumption is that passengers are required to disembark through the same units they board. This was assumed for maximum generality, as it allowed us to apply *OS-XLT* protocols to trains where passenger movement across units is impossible. But this assumption is unnecessarily restrictive for trains with an open interior that allows passengers to move across units. In this case, all the results given still hold, but actual performance may be even better. This occurs for three reasons: (i) the passenger loads may self-balance across units; (ii) some transferring passengers may be able to transfer while on board, eliminating their wait at a transfer station; and (iii) units without doors can be appended to both ends of each train—and never aligned—further increasing the train's passenger-holding capacity.

To illustrate improvement (ii), consider the two transfers described by the two passenger paths in Figure 5. Recall that their horizontal dotted segments denote a disembarkation at the transfer station, a walk along the platform and an embarkation at the next train. With an open interior design, however, these passengers could have walked inside the train while en route, avoiding the two (dis)embarkation moves and the intervening wait. For this to be possible, however, the O-D stations must be sufficiently far apart for the walk to be completed in time. Furthermore, passengers should be well informed so they can walk to the correct units. This could be achieved by displaying the train alignment bar-chart of Section 7 in every unit of the train. Unfortunately however, only transfers between the same train types can avail themselves of this opportunity—cross-type transfers such as those depicted on Figures 6 and 7 obviously cannot.

To better understand improvement (iii) (units-without-doors at both ends of a train) consider that with and open-interior the proposed end sections never have to be aligned with a platform, and yet they can be freely used by passengers going to any destination served by a nearby section. Obviously, these extra end sections can increase a train's length and maximum passenger occupancy. The improvement can be captured and optimized with the proposed analysis framework.

The second assumption was that the minimum time interval between $XLT$s is the same as between ordinary trains. This is not exactly true, however. The minimum safe headway between $XLT$s is actually slightly larger than for ordinary trains. To see this note that the headway relevant for capacity analysis is the front-to-front time separation between two successive trains, as measured by a stationary observer. The standard relevant for safety, however, is based on the rear-to-front time separation between trains—with due consideration of the maximum train speed. The difference between the capacity- and safety-separations is simply the time elapsed while a train goes by at cruising speed; i.e., $L/v$ [s] if the train is $L$ [m] long and cruises at speed $v$ [m/s]. Obviously then, if both train technologies are held to the same safety standards but $XLT$ trains are $\Delta L$ [m] longer than ordinary trains, the minimum $XLT$ headways will be $\Delta L/v$ [s] greater than those of ordinary trains; i.e., slightly larger as was pointed out. Fortunately, for typical applications this enlargement is small and only has a second order effect.[16]

---

[16] As point of reference, consider the longest (200 [m]) BART trains at the San Francisco Bay Area Embarcadero Station. This station is at the mouth of the Trans-Bay Tube (the busiest link in the BART system, which is at capacity) where trains travel at about 30 [m/s] (or 70 [miles/h]). The minimum headway is 157 [s] (or 23 [trains/h]). If trains at this station were extended by 70 [m] on account of the F/R protocol, headways would have

This paper has offered some 'proof-of-concept" ideas and a framework for systematically designing *XLT* operations, but the work is not finished. The following are some suggestions. While the formulation in Section 6 is general, the optimization problem described becomes unwieldy if all the variables in all seven tables are treated as decision variables. This is why static protocols were analyzed in more detail in Section 7, and why structured dynamic protocol families, such as the ones described in Sec. 4.2 are useful. Restricting our focus to particular protocol structures allows us to fix the variables in many of the tables, and optimize only the rest. In the author's view, research is needed both, to invent new protocol structures that would address particular problems, and to improve the design/optimization methods for any given structure. Methods could even be developed to make some of the decisions adaptively depending on the demand that is being experienced. These optimization methods should be developed for realistic scenarios involving time–dependent and possibly oversaturated conditions. A goal should be the development of *TOPS-XLT* planning software that can be put to practical use.

---

to be increased by $\Delta L/v = 2.5$ [s] to 159.5 [s]. Simple arithmetic shows that the F/T protocol's 33.3% improvement in passenger occupancy would actually be about 31.2%.